\documentclass[12pt]{article}
\newcommand{\drawsquare}[2]{\hbox{%
\rule{#2pt}{#1pt}\hskip-#2pt
\rule{#1pt}{#2pt}\hskip-#1pt
\rule[#1pt]{#1pt}{#2pt}}\rule[#1pt]{#2pt}{#2pt}\hskip-#2pt
\rule{#2pt}{#1pt}}

\newcommand{\Yfund}{\raisebox{-.5pt}{\drawsquare{6.5}{0.4}}}
\newcommand{\Ysymm}{\raisebox{-.5pt}{\drawsquare{6.5}{0.4}}\hskip-0.4pt%
        \raisebox{-.5pt}{\drawsquare{6.5}{0.4}}}
\newcommand{\Ythrees}{\raisebox{-.5pt}{\drawsquare{6.5}{0.4}}\hskip-0.4pt%
          \raisebox{-.5pt}{\drawsquare{6.5}{0.4}}\hskip-0.4pt%
          \raisebox{-.5pt}{\drawsquare{6.5}{0.4}}}
\newcommand{\Yfours}{\raisebox{-.5pt}{\drawsquare{6.5}{0.4}}\hskip-0.4pt%
          \raisebox{-.5pt}{\drawsquare{6.5}{0.4}}\hskip-0.4pt%
          \raisebox{-.5pt}{\drawsquare{6.5}{0.4}}\hskip-0.4pt%
          \raisebox{-.5pt}{\drawsquare{6.5}{0.4}}}
\newcommand{\Yasymm}{\raisebox{-3.5pt}{\drawsquare{6.5}{0.4}}\hskip-6.9pt%
        \raisebox{3pt}{\drawsquare{6.5}{0.4}}}
\newcommand{\Ythreea}{\raisebox{-3.5pt}{\drawsquare{6.5}{0.4}}\hskip-6.9pt%
        \raisebox{3pt}{\drawsquare{6.5}{0.4}}\hskip-6.9pt
        \raisebox{9.5pt}{\drawsquare{6.5}{0.4}}}

\newcommand{\jref}[4]{{\it #1} {\bf #2}, #3 (#4)}

\newcommand{\MPLA}[3]{\jref{Mod.\ Phys.\ Lett.}{A#1}{#2}{#3}}

\newcommand{\NPB}[3]{\jref{Nucl.\ Phys.}{B#1}{#2}{#3}}

\newcommand{\PLB}[3]{\jref{Phys.\ Lett.}{#1B}{#2}{#3}}

\newcommand{\PRD}[3]{\jref{Phys.\ Rev.}{D#1}{#2}{#3}}

\newcommand{\PRL}[3]{\jref{Phys.\ Rev.\ Lett.}{#1}{#2}{#3}}

\renewcommand{\theequation}{\thesection.\arabic{equation}}
\makeatletter
\def\vereq#1#2{\lower3pt\vbox{\baselineskip1.5pt \lineskip1.5pt
\ialign{$\m@th#1\hfill##\hfil$\crcr#2\crcr\sim\crcr}}}
\makeatother

\begin{document}

\begin{titlepage}
\begin{center}
hep-th/9801173     \hfill    LBNL-41320 \\
~{} \hfill UCB-PTH-98/07  \\
~{} \hfill UCSD/PTH 98-03 \\

\vskip .3in
 
{\Large \bf Classification of the  $N=1$ Seiberg-Witten Theories}

\vskip 0.3in

{\bf Csaba Cs\'aki$^{a,}$\footnote{Research Fellow, Miller Institute for 
Basic Research in Science.} and Witold Skiba$^b$}

\vskip 0.15in

{\bf $^a$} {\em Theoretical Physics Group\\
     Ernest Orlando Lawrence Berkeley National Laboratory\\
     University of California, Berkeley, CA 94720}

   and 

{\em Department of Physics\\
     University of California, Berkeley, CA 94720}

\vskip 0.1in

{\bf $^b$} {\em Department of Physics\\
University of California at San Diego, La Jolla, CA 92093}

\vskip 0.1in
 {\tt  csaki@thwk5.lbl.gov, skiba@einstein.ucsd.edu}

\end{center}

\vskip .25in

\begin{abstract}
We present a systematic study of $N=1$ supersymmetric gauge theories 
which are in the Coulomb phase. We show how to find all such theories
based on a simple gauge group and no tree-level superpotential. We find the
low-energy solution for the new theories in terms of a hyperelliptic
Seiberg-Witten curve. This work completes the study of
all $N=1$ supersymmetric gauge theories where the Dynkin index of the
matter fields equals the index of the adjoint ($\mu =G$), and consequently
all theories for which $\mu <G$.

\end{abstract}

\end{titlepage}

\newpage

\section{Introduction}
\setcounter{equation}{0}
The past four years have witnessed a tremendous progress in our understanding
of strongly coupled supersymmetric gauge theories. The number of theories
for which exact results have been established is ever growing since the 
initial breakthrough by Seiberg and Witten~\cite{SW}, who gave a complete 
solution of the  $N=2$ $SU(2)$ theory, and by Seiberg, who described the 
low-energy dynamics of $N=1$ supersymmetric QCD with varying number of
flavors~\cite{Seiberg}. 

    From these solutions the following basic picture emerges: there are 
six known phases of supersymmetric gauge theories: Higgs, confining,
abelian Coulomb, non-abelian Coulomb, free magnetic and infrared free phases. 
The actual phase of a given theory usually depends on the size of
its matter content. This size can be characterized by the relative value
of the Dynkin index $\mu$ of the chiral superfields compared to the value
of the Dynkin index $G$ of the vector superfields.  

In this paper we will focus on the $N=1$ theories which are in the
abelian Coulomb phase everywhere on their moduli spaces. 
We examine theories based on simple gauge groups and no tree-level
superpotential. We will argue that in order for such a theory to be in the
Coulomb phase, the theory has to satisfy the index condition $\mu =G$.
The essence of the argument can be summarized in the following:
one expects the low-energy solutions of such a theory to be given in terms
of an auxiliary Riemann surface, defined by a curve (which in most cases is
hyperelliptic). The classical curve is smoothed out by quantum corrections,
which are proportional to the dynamical scale $\Lambda$ of the theory.
We will show, that in the absence
of a tree-level superpotential the condition for $\Lambda$ to appear in the
curve is $\mu =G$. Theories with an adjoint chiral superfield (the pure
$N=2$ theories) satisfy this condition, and we will give a complete
list of other theories which do so as well. After restricting ourselves
to these theories it is easy to actually find all of those which are in the
Coulomb phase by checking the unbroken gauge group on a generic
point of the moduli space. This way we obtain a complete list of 
$N=1$ Seiberg-Witten theories based on simple gauge groups and no tree-level
superpotential. For these theories we determine the Seiberg-Witten
curves providing the solution for the low-energy effective gauge kinetic
couplings by flowing to theories for which the curve is already known.

This work completes the study of theories satisfying the index 
condition $\mu =G$. It has been known for a while, that confining theories
with a quantum deformed moduli space have to satisfy this index 
condition~\cite{s-conf,GN}. All such confining theories have been
systematically analyzed  in Refs.~\cite{s-conf,GN,GN2,Cho}.
We find, that if the matter content is in a faithful representation of
the gauge group, then a 
$\mu =G$ theory is confining with a quantum deformed moduli space. 
However, if the matter content is not in a faithful representation of the
gauge group, then the theory is in the Coulomb phase, and the low-energy
solution can be given in terms of a Seiberg-Witten curve. Therefore 
the low-energy dynamics of all theories with $\mu =G$ has now been 
determined. Since all theories with $\mu <G$ can be obtained from the
$\mu =G$ theories by adding mass terms for fields in vector-like
representations, it is a straightforward task to determine the low-energy
behavior of all $\mu <G$ theories as well.

The paper is organized as follows. In Section~\ref{sec:SW} we review the
basic properties of the Seiberg-Witten solutions and their applications
to $N=1$ theories. In Section~\ref{sec:arguments} we give our general
arguments which help us classify all $N=1$ Seiberg-Witten theories based
on simple groups and no tree-level superpotential. In 
Section~\ref{sec:solutions} we give the actual low-energy solutions of these
theories, and we conclude in Section~\ref{sec:Conclusions}. The derivation
of the curves for the new theories is explained in Appendices A and B.

\section{Review of the Seiberg-Witten Solution and its Application to
$N=1$ Theories\label{sec:SW}}
\setcounter{equation}{0}

Seiberg and Witten showed how to employ electric-magnetic duality to
obtain exact solutions to the low-energy dynamics of $N=2$ theories~\cite{SW}.
The original example of $SU(2)$ theory was subsequently generalized 
to other classical groups in Refs.~[7-17]. 
An alternative derivation of these solutions using the confining phase
superpotential method has been given in Ref.~\cite{confphase}, while the
connection to integrable systems has been described in~\cite{integrable,MW}.
The dynamics of the low-energy theory following from these solutions
has been analyzed in Refs.~\cite{sudynamics,dHoker}.
Below we review the basic features of these solutions of the $N=2$ theories,
and the application of the Seiberg-Witten methods to $N=1$ 
theories~\cite{phases}.

Since $N=2$ theories contain scalar fields in the adjoint representation,
their classical moduli space has a submanifold with unbroken $U(1)$ gauge
symmetries. In general, in addition to the Coulomb submanifold, there
is usually a subspace where the gauge group is completely broken.
The Higgs branch does not receive perturbative or non-perturbative quantum
corrections in $N=2$, while the Coulomb branch is affected by both
perturbative and non-perturbative effects. 

The low-energy Lagrangian of $N=2$ theories can be characterized in terms
of a single holomorphic function, the prepotential $\cal{F}$. In the
more interesting case of the Coulomb branch, the prepotential can be
computed in terms of the original ``electric'' fields and their dual
``magnetic'' degrees of freedom. It turns out that both kinds of variables
are necessary for a consistent description of the theory.

The classical pattern of symmetry breaking by a field in the adjoint
representation, $G\rightarrow U(1)^r$, where $r$ is the rank of the gauge
group, persists in the quantum theory everywhere on the Coulomb branch.
Classically, there are points with a larger unbroken subgroup, when 
the VEVs of the adjoint field happen to coincide. Due to quantum effects there
are no points of enhanced gauge symmetry, however, some states become
massless on certain submanifolds of the moduli space. These additional
massless states are indicated by singularities in the effective description.
Since these massless particles carry magnetic charges, magnetic variables
are more suitable for description of the theory near singularities.

Another indication that there are singularities on the moduli space
is the presence of nontrivial monodromies. At large expectation values,
the monodromy can be calculated using perturbation theory. The remaining
strong-coupling monodromies are guessed using symmetry arguments and
the requirement that the product of strong-coupling monodromies
has to equal the monodromy at infinity. Knowledge of all monodromies, that
is the singularity structure and the behavior at infinity is sufficient to
determine the full form of a holomorphic quantity, in this case 
the prepotential ${\cal F}$.

The low-energy solution is obtained by 
introducing an auxiliary Riemann surface of genus equal to the rank of the
gauge group. All holomorphic quantities can be computed as line integrals
over this surface. This surface will also encode the singularity
structure of the theory. In most cases, this surface turns out to be a
hyperelliptic surface, which can be defined by an equation of the form:
\begin{displaymath}
  y^2 = P(x,u_i,\Lambda),
\end{displaymath}
where $P$ is a polynomial in $x$, $u_i$ and $\Lambda$. The variables 
$x$ and $y$ are auxiliary parameters, while the $u_i$'s are the coordinates
on the moduli space and
$\Lambda$ is the dynamical scale of the theory. Therefore, finding
the exact low-energy action is equivalent to finding the polynomial
$P$. From the curve described by $P$ one can extract information about the
dynamics of the theory. In particular, the effective gauge coupling,
the metric on the moduli space and the spectrum of the BPS states can be 
calculated. 

For example, the simplest theory in the abelian Coulomb phase is $N=2$
$SU(2)$ with no hypermultiplets. In that case 
\begin{displaymath}
  y^2 = (x^2-u)^2 - 4\Lambda^4,
\end{displaymath}
where $u=\frac{1}{2} {\rm Tr} \Phi^2$ and $\Phi$ is the adjoint superfield.
This equation describes a torus, which is a surface of genus one.
Any VEV of an adjoint field can be rotated into the Cartan subalgebra,
which for $SU(2)$ means that $\langle \Phi \rangle = {\rm diag}(a,-a)$.
One needs also to introduce a dual scalar field $a_D$. The above curve
has three singularities, for $u=\pm \Lambda^2$ and $u=\infty$. From the
curve one can calculate the monodromies of the $(a_D,a)$ vector
around the singularities. The monodromies are elements of the duality
group, which acts on the vector $(a_D,a)$. The singularity at infinity is the 
perturbative singularity, while the two singularities at $u=\pm \Lambda^2$
occur at strong coupling. The non-perturbative singularities arise
because of a monopole or a dyon becoming massless at these points of the
moduli space. The charges of the massless fields
are the left eigenvectors of the respective monodromy  matrices.

  From the curve one can also compute $a$ and $a_D$ as functions of the
parameter of the moduli space $u$. These are determined as integrals
over the periods, $\gamma_{1,2}$, of the torus:
$a_D = \oint_{\gamma_1} \lambda$, $a = \oint_{\gamma_2} \lambda$,
where $\lambda$ is the so-called Seiberg-Witten differential. In the
case of $SU(2)$, $\lambda \propto \frac{x^2 \, dx}{y}$. The gauge coupling
$\tau=\frac{d a_D/d u}{d a/d u}=
    \frac{\partial^2 {\cal F}}{\partial a^2}$, from which
one can also establish the metric on the moduli space
$\frac{\partial {\cal F}}{\partial a}$.

Many features of the Seiberg-Witten solution of $N=2$ theories persist
in $N=1$ theories in the Coulomb phase. Intriligator and Seiberg pointed out
that the $U(1)$ gauge coupling, which is a holomorphic quantity,
can be described by the methods used in $N=2$~\cite{phases}. 
The dependence of the 
gauge coupling on the parameters of the moduli space can be found
once a curve describing the theory is established. The gauge-kinetic
term
\begin{displaymath}
{\cal L}=  \frac{1}{4\pi} {\rm Im} \int d^2 \theta \tau_{ij} W_\alpha^i 
W^{\alpha,j},
\end{displaymath}
where $\tau_{ij}$ is the effective gauge coupling matrix, whose eigenvalues 
are related to the effective gauge coupling and theta parameter of the k$^{th}$
$U(1)$ factor by
$ \tau_{k}=i \frac{4 \pi}{g_k^2} + \frac{\theta_k}{2 \pi}$.
The effective gauge coupling function $\tau_{ij}$
is not related by supersymmetry to the K\"ahler potential. Hence, no
information about the K\"ahler potential is provided by the $N=1$ solution.
Likewise, there is no central extension  of the $N=1$
supersymmetry algebra which would incorporate BPS particles.
However, one can learn about the charges of massless states associated with
singularities. The monodromy around a singularity still encodes the
information about the charge. These methods have been used in 
Refs.~\cite{phases,JoshDan,otherN=1} to obtain solutions to
several $N=1$ theories in the Coulomb phase or to Coulomb branches 
of $N=1$ theories with tree-level superpotential terms. 

\section{Necessary Criteria for Seiberg-Witten Theories\label{sec:arguments}}
\setcounter{equation}{0}

In this section we will show how to systematically find all $N=1$ 
supersymmetric  theories based on a simple gauge group and no tree-level 
superpotential, which are in the Coulomb phase. We will call such theories
the $N=1$ Seiberg-Witten theories. First we show that such theories have 
to satisfy the index condition $\mu =G$ mentioned in the introduction,
and then decide which $\mu =G$ theories are actually in the Coulomb phase.

\subsection{The Index Condition}

We have seen in the previous section that the effective gauge kinetic
function $\tau_{ij}$ can be identified with the period matrix 
of an auxiliary Riemann surface, usually a hyperelliptic curve
$y^2 = P(x,u_i,\Lambda)$. The curve obtained in the limit
$\Lambda \to 0$ is singular everywhere on the moduli space reflecting the
fact that turning off the gauge coupling will result in
additional massless gauge bosons independently of the VEVs of the scalars,
since there is no Higgs mechanism in the $\Lambda\to 0$ limit. 
This singularity must be smoothed out by effects proportional to
$\Lambda$,  except at a submanifold where
the singularity persists indicating the existence of additional massless 
states. The lesson which should be learned from this is that the dynamical
scale $\Lambda$ has to appear as a parameter of the full Seiberg-Witten
curve to smooth out the classical singularities. In the absence of a 
tree-level superpotential coupling this requirement will severely restrict
the matter content of the theory. 

We now discuss the constraint which arises from requiring that the dynamical
scale $\Lambda$ appears in the Seiberg-Witten curve. This curve must
respect all symmetries of the original theory. In particular, one can consider 
a $U(1)_R$ symmetry under which all fields $\Phi_i$ carry zero charge.
This symmetry is anomalous under the gauge group $G$, that is the
$G^2 U(1)_R$ anomaly is non-vanishing. One can however restore this
symmetry by promoting the dynamical scale $\Lambda$ to a background 
chiral superfield~\cite{ILS}. The reason for this is that an 
anomalous $U(1)$ rotation by an angle $\alpha$ acts like a shift 
\[ \theta \to \theta +\sum_i \mu_i q_i \alpha \]
on the $\theta$ parameter of the theory, where $q_i$ is the charge of the
i$^{th}$ fermion under the $U(1)$ transformation, and $\mu_i$ is the 
Dynkin index of the i$^{th}$ fermion\footnote{The Dynkin index is defined
by ${\rm Tr}\, T^aT^b=\mu \delta^{ab}$, where the $T$'s are the generators
of the gauge group in a given representation.}.
This means that the dynamical scale 
\[ \Lambda^{b}= \mu^be^{-\frac{8\pi^2}{g^2(\mu )}+i\theta}\]
of the theory has charge $\sum_i \mu_i q_i$ under the anomalous $U(1)$ 
symmetry, where $b$ is the coefficient of the one-loop beta function. 
 In the case of the $U(1)_R$ symmetry defined above, the 
fermions from the chiral superfield have charge $-1$, while the gauginos
have charge $+1$, thus the charge of $\Lambda^b$ is 
$G-\mu$, where $G$ is the Dynkin index of the adjoint representation,
and $\mu =\sum_i \mu_i$ is the sum of the Dynkin indices of the matter fields.
After assigning this charge $G-\mu $ to the dynamical scale, one can 
require that the Seiberg-Witten curve is invariant (or at least 
covariant) under this anomalous $U(1)_R$ symmetry as well. 

Let us now consider the $\Lambda \to 0$ limit again. In this limit we obtain
the classical curve $P_{cl}(y,x,u_i)=0$. Since we expect that the full
curve defined by $x$ and $y$ describes a Riemann surface, we expect that
$P_{cl}$ is not a homogeneous polynomial in $x$ and $y$ (which is true for
every known solution). However this implies that $x$ and $y$ have to have 
zero R-charge as well. But now considering the full curve 
$P(y,x,u_i,\Lambda )=0$ one can see that $\Lambda$ can appear in a non-trivial 
way in the curve only if its R-charge is zero as well, implying that
$\mu =G$. Note that all pure $N=2$ theories without matter fields satisfy 
this condition, however the $N=2$ theories with hypermultiplets do not.
The way these theories evade this constraint is that $N=2$ supersymmetry
automatically requires a tree-level superpotential coupling between the
hypermultiplets and the adjoint from the $N=2$ vector superfield, which
explicitly breaks the $U(1)_R$ symmetry used above. 

The $\mu =G$  condition is exactly the same as one gets for theories with a 
quantum deformed moduli space. The coincidence of these two conditions is
not very surprising, since in both cases we are requiring that $\Lambda $
appears in an equation involving the moduli in a non-trivial way. In the
case of the quantum deformed moduli space we require that a term proportional
to $\Lambda $ appears on the right hand side of a classical constraint,
while in the case of Seiberg-Witten theory we require that $\Lambda $ appears
as a non-trivial modification of the  classical Seiberg-Witten curve. 
 
One can argue for the necessity of the $\mu =G$ condition in a different way
as well. We are requiring that a Seiberg-Witten theory has unbroken
$U(1)$'s on the generic point in the moduli space. It is however 
known~\cite{DM}, that for $\mu >G$ the gauge group is completely broken, thus
the $\mu >G$ theories can not be in the Coulomb phase everywhere on the 
moduli space. If $\mu <G$, then a dynamically generated superpotential of the
form $W_{dyn} \propto \frac{1}{(\Pi_i \Phi_i^{\mu_i})^{\frac{2}{G-\mu}}}$
is allowed by all symmetries of the theory, and such a superpotential is
presumably generated either by instantons or by gaugino condensation. This 
superpotential pushes the fields to large expectation values and the theory
has no stable vacuum. Thus, we expect the $\mu <G$ theories to be generically 
in the Higgs phase, or at least have a Higgs branch with a runaway vacuum.
Thus we again conclude that the only possibility for a theory to be in the
Coulomb phase everywhere on the moduli space is when the index condition
$\mu =G$ is satisfied.

We have established that a necessary condition for Seiberg-Witten
theories is that the index condition $\mu =G$ be satisfied. This restricts the 
number of candidate theories considerably. In the next subsection we
show how to find the theories which are actually in the Coulomb phase.

\subsection{Flows}

We have seen in the previous section, that a necessary condition for a theory
to be in the Coulomb phase is that it satisfies the index condition
$\mu =G$. This requirement alone reduces the number of candidate 
theories to a finite set. In order to decide which of these theories is 
in the Coulomb phase we have to check whether there are unbroken 
$U(1)$'s on the generic point of the moduli space. 

However, in order to exclude a given candidate theory from being in the
Coulomb phase one doesn't always have to consider the most generic point on
the moduli space. It is enough to find a flow which leads to a theory which is
known not to be in the Coulomb phase. For example, consider the 
theory based on the exceptional group $E_7$ with three $56$ dimensional
representations. This theory satisfies the $\mu =G$ constraint. However, by
giving a VEV to one of the $\bf{5}\bf{6}$'s we get an $E_6$ theory with
$2\cdot({\bf 27}+{\bf \overline{27}})$. Giving an expectation value to the 
${\bf 27}$ breaks
$E_6$ to $F_4$, with the remaining field content being three $26$-dimensional
representations. Further breaking $F_4$ by a VEV of ${\bf 26}$ will give
$SO(9)$ with two spinors and three vectors, $2  \cdot{\bf 16}+3 \cdot{\bf 9}$.
However, this $SO(9)$ theory is known to be confining with a quantum 
modified constraint~\cite{s-conf}, and is therefore not in the Coulomb phase. 
This argument implies that the whole chain of theories, with matter content
such that $\mu =G$, given below is excluded from being in the Coulomb phase.
\begin{equation}
 E_7:\; [3\cdot{\bf 56}] \stackrel{\langle {\bf 56} 
  \rangle}{\rightarrow} 
   E_6:\; [2 \cdot{\bf 27}+2  \cdot{\bf \overline{27}}] 
   \stackrel{\langle {\bf 27} \rangle}{\rightarrow} 
   F_4: [3  \cdot{\bf 26}] \stackrel{\langle {\bf 26} \rangle}{\rightarrow}
   SO(9)\; [2 \cdot{\bf 16}+3  \cdot{\bf 9}]
\end{equation}
Similarly, one could consider an $E_6$ theory with $n\; \cdot{\bf 27}+
(4-n)\; \cdot{\bf \overline{27}}$, where $n=0,1,2,3,4$. All of these 
theories 
will also flow to $F_4$ with $3\; \cdot{\bf 26}$, so they are not in the
Coulomb phase either. Indeed, it has been shown recently in
Refs.~\cite{GN2,Cho}, that all of the above theories based on
exceptional groups satisfying $\mu =G$ are confining with a quantum
deformed moduli space. 

In Tables 1-4 we list all theories that satisfy the $\mu =G$ constraint
and give the phase of the given theory. The first column gives the
gauge group, the second column the field content and the third column the 
phase of the theory. Finally, the fourth column  contains a reference
to where the actual low-energy solution of the given theory can be
found. One can see from Tables 1-4 that finding the Seiberg-Witten curves
for the remaining $N=1$ theories in the Coulomb phase completes the
study of all $\mu =G$ theories.

There are only two possibilities for the phase of the $\mu =G$ theories:
confining phase or Coulomb phase. The confining theories
all have a low-energy description in terms of composite gauge invariants
satisfying a quantum-modified version of the classical constraints
(the $SU,Sp$ and $SO$ theories in this class have been analyzed in 
Refs.~\cite{s-conf,GN,GN2}, while the theories based on exceptional groups in 
Refs.~\cite{GN2,Cho}.)
This solution is valid everywhere on the moduli space, and there is no phase
boundary between the Higgs and the confining phases. This is possible,
because the massless fermions are in a faithful representation of the 
gauge group, therefore any external source can be screened by the massless 
fields. One can also explicitly check that every $\mu =G$ theory
with chiral superfields in a faithful representation breaks
the gauge group completely at generic points on the moduli space.
Since at large expectation values the theory can be described entirely
in terms of gauge singlet fields and there is no invariant distinction
between the Higgs and the confining phase, one indeed expects confinement
at strong coupling. Contrary, if chiral superfields  are not in a
faithful representation of the gauge group, some external sources can not
be screened by the massless quarks, and there can be points on the moduli
space where additional massless fields appear. Indeed, we find that in every
case where the matter fields are not in a faithful representation, at large
expectation values there are unbroken $U(1)$ gauge factors and that the
low-energy theory is in the Coulomb phase. One can see from Tables 1-4 that
the only theories which are in the Coulomb phase besides the pure $N=2$
theories (which are $N=1$ theories with a chiral superfield in the adjoint
representation) are $SO(N)$ with $N-2$ vectors, $SU(6)$ with $2\, \Ythreea$
and $Sp(6)$ with $2\, \Yasymm$. The other two theories in the Coulomb phase
belong to the $SO(N)$ series with $N-2$ vectors, since $SU(4)$ with
$4\; \Yasymm$ is equivalent to $SO(6)$ with four vectors and $Sp(4)$ with
$3\; \Yasymm$ is equivalent to $SO(5)$ with three vectors. 

One can easily see that these $N=1$ theories are indeed in the Coulomb phase 
by considering the following flows:
\begin{eqnarray}
\label{flows}
&& SO(N): \; (N-2) \Yfund  \rightarrow
 SU(2)\times SU(2): \; 2\; 
(\Yfund ,\Yfund) \nonumber \\
&& SU(6): \; 2\; \Ythreea \rightarrow SU(3)\times SU(3): \; (\Yfund ,
\overline{\Yfund})+(\overline{\Yfund},\Yfund ) \nonumber \\
&& Sp(6): \; 2\; \Yasymm \rightarrow SU(2)\times SU(2) \times SU(2): \;
(\Yfund ,\Yfund ,1)+(1,\Yfund ,\Yfund )+(\Yfund ,1,\Yfund ).
\end{eqnarray}
In the above flows the following fields have to get expectation values:
in the $SO(N)$ theories $N-4$ vectors, in the $SU(6)$ theory one $\Ythreea$ and
in the $Sp(6)$ theory one $\Yasymm$. Such product group theories have been
shown to be in the Coulomb phase in Refs.~\cite{phases,JoshDan}. The $SO(N)$
series in (\ref{flows}) has been described in Ref.~\cite{SO}, while the
remaining flows will enable us to find the Seiberg-Witten curves for the
$SU(6)$ and $Sp(6)$ theories. The results are summarized in the next section,
while the detailed derivations are presented in Appendices A and B.

\begin{table}
\begin{center}
\[
\begin{array}{|c|c|c|c|} \hline
SU(N)   &   N (\Yfund + \overline{\Yfund})   & \mbox{confining} 
& \cite{Seiberg}   \\
SU(N)   &  \Yasymm + (N-1)\, \overline{\Yfund} + 3\, \Yfund 
                                              & \mbox{confining} 
& \cite{SUantisymm} \\
SU(N) & \Yasymm + \overline{\Yasymm} + 2 (\Yfund +
\overline{\Yfund})     		& \mbox{confining} 
& \cite{s-conf}   \\
SU(N) & Adj   & \mbox{Coulomb phase} & \cite{SW,AF,Klemm} \\ 
SU(4) & 3\, \Yasymm +(\Yfund + \overline{\Yfund}) &  \mbox{confining} 
& \cite{s-conf,GN} \\
SU(4) &  4\, \Yasymm  &   \mbox{Coulomb phase} & \cite{SO} \\ 
SU(5) &  2\, \Yasymm +  \Yfund + 3\, \overline{\Yfund} 	
					& \mbox{confining}  & \cite{s-conf} \\
SU(5) & 2\, \Yasymm + \overline{\Yasymm} +  
	\overline{\Yfund} & \mbox{confining} &  \cite{s-conf} \\ 
SU(6) & 2\, \Yasymm + 4\, \overline{\Yfund} & \mbox{confining} & \cite{s-conf}
 \\
SU(6) & \Ythreea + 3 (\Yfund + \overline{\Yfund}) & \mbox{confining} 
& \cite{s-conf} \\
SU(6) & \Ythreea + \Yasymm +  2\, \overline{\Yfund} 
& \mbox{confining} & \cite{GN} \\
SU(6) &  2\, \Ythreea  & \mbox{Coulomb phase} & \mbox{Appendix B} \\  
SU(7) &  \Ythreea + 4\, \overline{\Yfund} + 2\, 
		\Yfund  & \mbox{confining} & \cite{GN} \\  \hline
\end{array}\]
\end{center}
\caption{The $SU$ theories satisfying the index constraint $\mu =G$. The 
first column gives the gauge group, the second column the field content and
the third column gives the phase of the low-energy theory. The last column
gives a reference to where the low-energy solution of the
given theory can be found.}
\label{SUtable}
\end{table}

\begin{table}                     
\begin{center}
\[
\begin{array}{|c|c|c|c|} \hline
Sp(2N) & (2N+2)\, \Yfund & \mbox{confining} & \cite{IntPoul} \\ 
Sp(2N) & \Yasymm +4\, \Yfund  & \mbox{confining} & \cite{Spanti} \\
Sp(2N) & \Ysymm =Adj  & \mbox{Coulomb phase} & \cite{Sp,MW} \\  
Sp(4)  & 2\, \Yasymm +2\, \Yfund  & \mbox{confining} & \cite{GN} \\
Sp(4) & 3\, \Yasymm   &  \mbox{Coulomb phase}  & \cite{SO} \\  
Sp(6) & 2\, \Yasymm   &  \mbox{Coulomb phase}  & \mbox{Appendix A} \\  
Sp(6) & \Ythreea +3\,\Yfund  & \mbox{confining} & \cite{GN} \\
 \hline
\end{array} \]
\end{center}
\caption{The $Sp$ theories satisfying the index constraint $\mu =G$. The 
first column gives the gauge group, the second column the field content and
the third column gives the phase of the low-energy theory. The last column
gives a reference to where the low-energy solution of the
given theory can be found.}
\label{SPtable}
\end{table}

\begin{table}                     
\begin{center}
\[
\begin{array}{|c|c|c|c|} \hline
SO(N) & \Yasymm =Adj & \mbox{Coulomb phase} & \cite{SO2N,SO2N+1,MW}\\
SO(N) & (0,N-2) & \mbox{Coulomb phase} & \cite{SO} \\
SO(7) & (1,4) &  \mbox{confining} & \cite{G2}\\
SO(7) & (2,3) & \mbox{confining} & \cite{PoulStrassler}\\
SO(7) & (3,2) &  \mbox{confining} & \cite{s-conf} \\
SO(7) & (4,1) &   \mbox{confining} & \cite{s-conf} \\
SO(7) & (5,0) & \mbox{confining} & \cite{s-conf} \\
SO(8) & (3,2,1) &  \mbox{confining} & \cite{s-conf} \\
SO(8) & (2,2,2) &  \mbox{confining} & \cite{s-conf} \\
SO(8) & (4,2,0) &  \mbox{confining} & \cite{s-conf} \\
SO(8) & (4,1,1) &  \mbox{confining} & \cite{PoulStrassler} \\
SO(8) & (3,3,0) &  \mbox{confining} & \cite{s-conf}\\
SO(8) & (5,1,0) &  \mbox{confining}  & \cite{PoulStrassler} \\
SO(9) & (1,5) &  \mbox{confining} & \cite{PoulStrassler} \\
SO(9) & (2,3) &  \mbox{confining} & \cite{s-conf} \\
SO(9) & (3,1) &  \mbox{confining} & \cite{s-conf}\\
SO(10) & (1,0,6) &  \mbox{confining} & \cite{PoulStrassler}\\
SO(10) & (2,2,0) &  \mbox{confining} & \cite{s-conf} \\
SO(10) & (1,1,4) &  \mbox{confining} & \cite{s-conf}\\
SO(10) & (2,1,2) & \mbox{confining} & \cite{s-conf}\\
SO(10) & (3,1,0) &  \mbox{confining} & \cite{s-conf}\\
SO(10) & (2,0,4) &  \mbox{confining} & \cite{s-conf}\\
SO(10) & (3,0,2) &  \mbox{confining} & \cite{s-conf}\\
SO(10) & (4,0,0) &  \mbox{confining} & \cite{s-conf}\\
SO(11) & (2,1) &  \mbox{confining} & \cite{s-conf}\\
SO(11) & (1,5) &  \mbox{confining} & \cite{s-conf}\\
SO(12) & (1,1,2) &  \mbox{confining} & \cite{s-conf}\\
SO(12) & (2,0,2) &  \mbox{confining} & \cite{s-conf}\\
SO(12) & (1,0,6) &  \mbox{confining} & \cite{s-conf}\\
SO(13) & (1,3) &  \mbox{confining} & \cite{s-conf}\\
SO(14) & (1,0,4) &  \mbox{confining} & \cite{s-conf}\\ \hline
\end{array} \]
\end{center}
\caption{The $SO$ theories satisfying $\mu =G$. The 
first column gives the gauge group, the second one the field content and
the third one the phase of the low-energy theory. The last column
gives a reference to where the low-energy solution of the given theory
is described. We use the following notation for the field content: for an
$SO(N)$ group for $N$ odd $(s,v)$ denotes the number of spinors and the
number of vectors, while for $N$ even $(s,s',v)$ denotes the number of
the two inequivalent spinor  representations and the number of vectors. 
We do not distinguish between $SO(N)$ and its covering group $Spin(N)$.}
\label{SOtable}
\end{table}

\begin{table}                     
\begin{center}
\[
\begin{array}{|c|c|c|c|} \hline
G_2 & 4\, \times {\bf 7} & \mbox{confining} & \cite{G2} \\
G_2 & {\bf 14} = Adj & \mbox{Coulomb phase} & \cite{G2Coulomb,MW,G2F4E6} \\
F_4 & 3\, \times {\bf 26} & \mbox{confining} & \cite{GN2,Cho} \\
F_4 & {\bf 52} = Adj & \mbox{Coulomb phase} & \cite{MW,exceptional,G2F4E6} \\
E_6 & n \, \times {\bf 27} +(4-n)\, \times {\bf \overline{27}}, 
& \mbox{confining} & \cite{GN2,Cho} \\
E_6 & {\bf 78} = Adj & \mbox{Coulomb phase} & \cite{MW,exceptional,G2F4E6} \\
E_7 & 3\, \times {\bf 56} & \mbox{confining} & \cite{GN2,Cho} \\
E_7 & {\bf 133} =Adj & \mbox{Coulomb phase} & \cite{MW,exceptional} \\
E_8 & {\bf 248} =Adj & \mbox{Coulomb phase} & \cite{MW,exceptional} \\ \hline
\end{array} \]
\end{center}
\caption{The theories based on exceptional groups satisfying the index
constraint $\mu =G$. The first column gives the gauge group, the second column
the field content and the third column gives the phase of the low-energy theory.
The last column gives a reference to where the low-energy solution of the
given theory can be found.}
\end{table}

\section{The $N=1$ Seiberg-Witten Theories\label{sec:solutions}}
\setcounter{equation}{0}

In this section we give the low-energy solution of the $N=1$ Seiberg-Witten
theories. Since the solutions to the theories with one adjoint (the 
pure $N=2$ theories) are well-known, we refer the reader interested 
in these theories to the references given in the last column of Tables 1-4.
Below, we give only the solution to the $N=1$ theories considered
in (\ref{flows}), and do not list all $N=2$ examples. For these theories,
we first give the high-energy field
content and the unbroken global symmetries, then list the low-energy degrees 
of freedom which satisfy the 't Hooft anomaly matching conditions. 
Finally we give the curve for every theory which provides the low-energy
solution for the effective $U(1)$ gauge coupling.

\subsection{$SU(6)$ with $2\, \protect\Ythreea$}
\[
\begin{array}{c|c|ccc}
& SU(6) & SU(2) & U(1)_R & Z_{12} \\ \hline
A & \Ythreea & \Yfund & 0 & 1 \\ \hline
S=A^2 & 1 & 1 & 0 & 2 \\
T=A^4 & 1 & \Yfours & 0 & 4 \\
U=A^6 & 1 & 1 & 0 & 6 \\
\end{array}\]

The Seiberg-Witten curve is
\begin{eqnarray}
   y^2&=& \Big[ x^3-  \Big( \frac{9}{2} (T^2) +\frac{11}{3} S^4 -S U \Big)  x
         - 2^7 \Lambda^{12}- \nonumber \\
     &&    \Big( 2 S^3 U -\frac{107}{27} S^6 - \frac{1}{4} U^2 
          - \frac{3}{2} S^2 (T^2) + 9 (T^3) \Big) \Big]^2 - 2^{14} \Lambda^{24}.
\end{eqnarray}
A detailed description of this theory is given in Appendix B.

\subsection{$SO(N)$ with $(N-2)\, \protect\Yfund$ \protect\cite{SO}}

\[ 
\begin{array}{c|c|ccc}
& SO(N) & SU(N-2) & U(1)_R & Z_{2N-4} \\ \hline
Q & \Yfund & \Yfund & 0 & 1 \\ \hline
M=Q^2 & 1 & \Ysymm & 0 & 2 \\
\end{array}
\]

Note that in this example the anomaly matching conditions at the origin are 
not satisfied by the meson field $M$ itself. The reason is that a number
of monopoles become massless exactly at the origin, and their contribution
to the anomalies has to be taken into account. For details see 
Refs.~\cite{SO,discrete}. The Seiberg-Witten curve for this theory is:

\[
y^2=(x^2-({\rm det} M-8\Lambda^{2N-4}))^2-64\Lambda^{4N-8}.
\]

\subsection{$Sp(6)$ with $2\, \protect\Yasymm$}

\[
 \begin{array}{c|c|ccc}
       & Sp(6) & SU(2) & U(1)_R & Z_8  \\ \hline
  A_i & \Yasymm & \Yfund & 0  & 1 \\ \hline
  S_{ij}={\rm Tr}(J A_i J A_j) & & \Ysymm & 0 & 2 \\
  T_{ijk}={\rm Tr}(J A_i J A_j J A_k) & & \Ythrees & 0 & 3 \\
  U={\rm Tr}(J A_1 J A_2 J A_1 J A_2) & & 1 & 0 & 4
 \end{array}
\]

The Seiberg-Witten curve is given by
\begin{eqnarray}
 y^2 &=& \Big[ x^2-(- 72 \Lambda^8 \Big( (S^2) + 12 U \Big)  + 
               \Big( -\frac{1}{2} (T^4) + 24  U^3 +
     \frac{3}{2} U^2 (S^2)+ \nonumber \\
   &&   + 3 U (S T^2) + \frac{1}{4} (S^3 T^2) \Big) \Big]^2
         - 768 \Lambda^{24}.
\end{eqnarray}
A detailed description of this theory is given in Appendix A.

\section{Conclusions\label{sec:Conclusions}}

We have studied $N=1$ supersymmetric gauge theories which are in the 
Coulomb phase on the entire moduli space. We have shown that theories
based on a simple gauge group
and no tree-level superpotential must satisfy the index condition
$\mu =G$, which is exactly the same as for theories with a quantum 
deformed moduli space. One can find the theories which are actually 
in the Coulomb phase by studying the flows of the theory. It turns out that
all $\mu =G$ theories are either confining with a quantum deformed moduli
space if the matter content is in the faithful representation of the
gauge group or in the Coulomb phase if the matter fields are not in a 
faithful representation. The Seiberg-Witten curves for the new theories 
in the Coulomb phase can be found by studying the flows to the product
group theories of Ref.~\cite{JoshDan}. This work, together with the results
on confining theories with quantum-deformed moduli spaces, completes
the solutions to all $N=1$ theories with $\mu\leq G$.

\section*{Acknowledgements}

We are grateful to Daniel Freedman 
for comments on the manuscript. C.C. is a Research
Fellow at the Miller Institute for Basic Research in Science. C.C.
has been supported in part by the U.S. Department of Energy under
contract DE-AC03-76SF00098 and in part by the National Science Foundation
under grant PHY-95-14797. W.S. was supported by the U.S. Department of Energy
under contract DE-FG03-97ER405046.

\section*{Appendix A \hspace*{0.25cm} $Sp(6)$ with $2\, \Yasymm$}
\setcounter{equation}{0}
\renewcommand{\theequation}{A.\arabic{equation}}

In this appendix we outline the derivation of the Seiberg-Witten
curve for the $Sp(6)$ theory with $2\, \Yasymm$. As we already mentioned
in Section~\ref{sec:arguments}, giving a VEV to one of antisymmetric tensors
breaks $Sp(6)$ to $SU(2)^3$ with precisely the field content that was
considered in Ref.~\cite{JoshDan}. The curve describing the $Sp(6)$ theory
must therefore reduce to the curve for $SU(2)^3$ in the limit of large VEVs.
It turns out that considering this limit is sufficient for determining the
complete curve.

Lets us first describe the global symmetries of the theory and gauge invariant
operators parameterizing the moduli space.
\begin{equation}
 \begin{array}{c|c|ccc}
       & Sp(6) & SU(2) & U(1)_R & Z_8  \\ \hline
  A_i & \Yasymm & \Yfund & 0  & 1 \\ \hline
  S_{ij}={\rm Tr}(J A_i J A_j) & & \Ysymm & 0 & 2 \\
  T_{ijk}={\rm Tr}(J A_i J A_j J A_k) & & \Ythrees & 0 & 3 \\
  U={\rm Tr}(J A_1 J A_2 J A_1 J A_2) & & 1 & 0 & 4
 \end{array},
\end{equation}
where $J$ is the two-index antisymmetric invariant tensor of $Sp$. Indeed, with this
choice of operators the 't Hooft anomaly matching conditions (including the
discrete anomaly matching conditions of Ref.~\cite{discrete})
are satisfied at the origin, once a $U(1)$ vector field is included in the
low-energy spectrum.

We give a VEV
to $A_1$ of the form $\langle A_1 \rangle=i v\, {\rm diag}(\sigma_2, \omega \sigma_2,
\omega^2 \sigma_2)$, where $\omega=\exp{(2 \pi i/3)}$. The remaining field, $A_2$,
decomposes under $SU(2)\times SU(2) \times SU(2)$ as
\begin{displaymath}
  A_2 \rightarrow Q_1(\Yfund,\Yfund,1)+Q_2(1,\Yfund,\Yfund)+ Q_3(\Yfund,1,\Yfund) +
     s_{1,2}(1,1,1).
\end{displaymath}
Next, we express the invariants of the $SU(2)^3$ theory~\cite{JoshDan},
that is $M_i=Q_i^2$ and $T=Q_1 Q_2 Q_3$ in terms of the $Sp(6)$ gauge invariants
keeping only leading terms in $1/v$. The curve for $SU(2)^3$
\begin{displaymath}
  y^2=[x^2-(\Lambda_1^4 M_2 + \Lambda_2^4 M_3 + \Lambda_3^4 M_1 + T^2 - M_1 M_2 M_3)]^2
       - 4 \Lambda_1^4 \Lambda_2^4 \Lambda_3^4 
\end{displaymath}
contains two combinations of invariants. We will now write the two combinations,
$\Lambda_1^4 M_2 + \Lambda_2^4 M_3 + \Lambda_3^4 M_1 $ and $T^2 - M_1 M_2 M_3$,
in terms of $Sp(6)$ invariants, where $\Lambda_i$ is the characteristic scale of 
the $i$-th $SU(2)$ factor. With the above choice of the VEV for $A_1$, the scale
matching relations between the $Sp(6)$ scale, $\Lambda$, and the scales of the
$SU(2)$ factors are 
\begin{displaymath}
  \Lambda_1^4  = \frac{\Lambda^8}{v^4 (1-\omega^2) (1-\omega^4)}, \; \;
  \Lambda_2^4  = \frac{\Lambda^8}{v^4 (\omega^2-1) (\omega^2-\omega^4)}, \; \;
  \Lambda_3^4  = \frac{\Lambda^8}{v^4 (\omega^4-1) (\omega^4-\omega^2)}. 
\end{displaymath}

Since a curve can depend only on the field combinations invariant under
global symmetries, we need to introduce invariants under the global
$SU(2)$ symmetry, which will be the variables of the full $Sp(6)$ curve.
The connection between the global $SU(2)$ invariants and the parameters of
the $SU(2)^3$ curve are
\begin{eqnarray}
  && \Lambda_1^4 \Lambda_2^4 \Lambda_3^4=\frac{1}{27 v^{12}} \Lambda^{24}, \nonumber \\
  && \Lambda_1^4 M_2 + \Lambda_2^4 M_3 + \Lambda_3^4 M_1=- \frac{1}{72 v^6} \Lambda^8
      \Big( (S^2) + 12 U \Big), \nonumber \\
  && T^2 - M_1 M_2 M_3= \frac{1}{5184 v^6} \Big( -\frac{1}{2} (T^4) + 24  U^3 +
     \frac{3}{2} U^2 (S^2)+3 U (S T^2) + \frac{1}{4} (S^3 T^2) \Big), \nonumber
\end{eqnarray}
where the $SU(2)$ contractions are defined as follows 
\begin{eqnarray}
  (S^2)&=&S_{i_1 i_2} S_{j_1 j_2} \epsilon^{i_1 j_1} \epsilon^{i_2 j_2} \nonumber \\
  (T^4)&=&T_{i_1 i_2 i_3} T_{j_1 j_2 j_3} T_{k_1 k_2 k_3} T_{l_1 l_2 l_3}
           \epsilon^{i_1 j_1} \epsilon^{i_2 j_2} \epsilon^{k_1 l_1} \epsilon^{k_2 l_2}
           \epsilon^{i_3 k_3} \epsilon^{j_3 l_3}, \nonumber \\
  (S T^2)&=&T_{i_1 i_2 i_3} T_{j_1 j_2 j_3} S_{k_1 k_2} \epsilon^{i_1 j_1}
             \epsilon^{i_2 j_2} \epsilon^{i_3 k_1} \epsilon^{j_3 k_2}, \nonumber \\
  (S^3 T^2)&=&T_{i_1 i_2 i_3} T_{j_1 j_2 j_3} S_{k_1 k_2} S_{l_1 l_2} S_{m_1 m_2} \times
               \nonumber \\
           &&  ( \epsilon^{i_1 j_1} \epsilon^{i_2 j_2} \epsilon^{i_3 k_1}
               \epsilon^{k_2 l_1} \epsilon^{l_2 m_1} \epsilon^{m_2 j_3}
               -\frac{1}{3} 
                \epsilon^{i_1 k_1} \epsilon^{i_2 l_1} \epsilon^{i_3 m_1}
               \epsilon^{j_1 k_2} \epsilon^{j_2 l_2} \epsilon^{j_3 m_2}). \nonumber 
\end{eqnarray}

Finally, we obtain the curve for $Sp(6)$ theory in the large VEV limit by 
substituting
the above expression into the $SU(2)^3$ curve:
\begin{eqnarray}
  y^2 &=& \Big[ x^2-(- \frac{1}{72 v^6} \Big( (S^2) + 12 U \Big) \Lambda^8 + 
             \frac{1}{5184 v^6} [
         \Big( -\frac{1}{2} (T^4) + 24  U^3 +
     \frac{3}{2} U^2 (S^2)+ \nonumber \\
   &&   3 U (S T^2) + \frac{1}{4} (S^3 T^2) \Big) \Big]^2
         - \frac{4}{27 v^{12}} \Lambda^{24}.
\end{eqnarray}
After rescaling $x\rightarrow x/(72 v^3)$ and $y\rightarrow y/(5184 v^6)$ we obtain:
\begin{eqnarray}
\label{Sp6}
 y^2 &=& \Big[ x^2-(- 72 \Lambda^8 \Big( (S^2) + 12 U \Big)  + 
               \Big( -\frac{1}{2} (T^4) + 24  U^3 +
     \frac{3}{2} U^2 (S^2)+ \nonumber \\
   &&   3 U (S T^2) + \frac{1}{4} (S^3 T^2) \Big) \Big]^2
         - 768 \Lambda^{24}.
\end{eqnarray}
It is easy to check that symmetries prohibit modifications to this form of the curve.
Since only $SU(2)$ singlets can appear in the curve, any other allowed term would be of
the same order in $1/v$ and therefore a coefficient of any such terms must be zero. Thus (\ref{Sp6}) is the final form of the full $Sp(6)$ curve.

\section*{Appendix B  \hspace*{0.25cm} $SU(6)$ with $2\, \Ythreea$}
\setcounter{equation}{0}
\renewcommand{\theequation}{B.\arabic{equation}}

We now present the derivation of the curve for the $SU(6)$ theory with $2 \, \Ythreea$.
Similarly to the derivation of the $Sp(6)$ case, we give a VEV to one of the tensors,
which breaks $SU(6)$ to $SU(3)\times SU(3)$ with two bifundamental fields, which is the
theory considered in Ref.~\cite{JoshDan}. The field content, global symmetries and independent
gauge invariants are defined below:
\begin{equation}
 \begin{array}{c|c|ccc}
       & SU(6) & SU(2) & U(1)_R & Z_{12}  \\ \hline
  A_i & \Ythreea & \Yfund & 0  & 1 \\ \hline
  S=A_1 A_2 & & 1 & 0 & 2 \\
  T_{ijkl}=A_i A_j A_k A_l & & \Yfours & 0 & 4 \\
  U=A_1^3 A_2^3 & & 1 & 0 & 6
 \end{array}.
\end{equation}
The anomaly matching is satisfied for the low-energy spectrum including the gauge invariants
and two $U(1)$ vector multiplets. The $SU(6)$ gauge contractions for the invariants $S$, $T$
and $U$ are given by
\begin{eqnarray}
  S&=&\frac{1}{6^2} A_1^{\alpha_1 \alpha_2 \alpha_3} A_2^{\beta_1 \beta_2 \beta_3}
      \epsilon_{\alpha_1 \alpha_2 \alpha_3 \beta_1 \beta_2 \beta_3}, \nonumber \\
  T_{ijkl} &=&\frac{1}{6^4}\left[A_i^{\alpha_1 \alpha_2 \alpha_3} A_j^{\beta_1 \beta_2 \beta_3}
                A_k^{\gamma_1 \gamma_2 \gamma_3} A_l^{\delta_1 \delta_2 \delta_3}
                \epsilon_{\alpha_1 \alpha_2 \alpha_3 \beta_1 \beta_2 \gamma_1}
                \epsilon_{\beta_3 \gamma_2 \gamma_3 \delta_1 \delta_2 \delta_3} - 
                \frac{S^2}{9} (\epsilon_{i j} \epsilon_{k l} + \epsilon_{i k} 
                   \epsilon_{j l}) \right], \nonumber \\
  U&=&\frac{1}{144}A_1^{\alpha_1 \alpha_2 \alpha_3} A_1^{\beta_1 \beta_2 \beta_3}
      A_1^{\gamma_1 \gamma_2 \gamma_3} A_2^{\delta_1 \delta_2 \delta_3}
      A_2^{\zeta_1 \zeta_2 \zeta_3} A_2^{\eta_1 \eta_2 \eta_3}
      \epsilon_{\alpha_1 \alpha_2 \alpha_3 \beta_1 \beta_2 \gamma_1}
      \epsilon_{\beta_3 \gamma_2 \gamma_3 \delta_1 \delta_2 \zeta_1}
      \epsilon_{\delta_3 \zeta_2 \zeta_3 \eta_1 \eta_2 \eta_3}. \nonumber
\end{eqnarray}
Note the additional subtlety in the definition of $T$. $T$ is a four-index symmetric tensor under
the global $SU(2)$. However, the gauge contraction defined above does not yield an irreducible
tensor of $SU(2)$. This contraction also contains an $SU(2)$ singlet piece, which needs to
be subtracted. 

We give a VEV to $A_1$ of the form $\langle A_{123} \rangle=\langle A_{456} \rangle=v$
with zero VEVs for all other independent components. The field $A_2$ decomposes
under $SU(3)\times SU(3)$ as
\begin{displaymath}
  A_2\rightarrow Q_1(\Yfund,\overline{\Yfund}) + Q_2(\overline{\Yfund},\Yfund) +s_{1,2}(1,1,1).
\end{displaymath}
The $SU(3)^2$ theory has four invariants~\cite{JoshDan}, $B_i=\det Q_i$ and
$T_i={\rm Tr}(Q_1 Q_2)^i$, $i=1,2$, which we now express in terms of $SU(6)$ invariants.
The $SU(3)\times SU(3)$ curve is
\begin{displaymath}
 y^2=(x^3-u_2 x -u_3 -\Lambda_1^6 -\Lambda_2^6)^2 - 4 \Lambda_1^6 \Lambda_2^6,
\end{displaymath}
where $\Lambda_{1,2}$ are the scales of the two $SU(3)$ gauge groups, 
while
$u_2=\frac{1}{2} (T_2 -\frac{1}{3} T_1^2)$
and $u_3=\frac{1}{3} (3 B_1 B_2 +\frac{1}{2} T_2 T_1 -\frac{5}{18}T_1^3)$.
In terms of $SU(6)$ invariants we have 
\begin{eqnarray}
  u_2&=&\frac{1}{16 v^4} \left( \frac{9}{2} (T^2) +\frac{11}{3} S^4 -S U \right) \nonumber \\
  u_3&=&\frac{1}{64 v^6} \left( 2 S^3 U -\frac{107}{27} S^6 - \frac{1}{4} U^2 
      - \frac{3}{2} S^2 (T^2) + 9 (T^3) \right), \nonumber
\end{eqnarray}
where the invariants under the global $SU(2)$ are defined as follows:
\begin{eqnarray}
  (T^2)& =&T_{i_1 i_2 i_3 i_4} T_{j_1 j_2 j_3 j_4} \epsilon^{i_1 j_1} \epsilon^{i_2 j_2}
          \epsilon^{i_3 j_3} \epsilon^{i_4 j_4}, \nonumber \\
  (T^3)&=& T_{i_1 i_2 i_3 i_4} T_{j_1 j_2 j_3 j_4} T_{k_1 k_2 k_3 k_4} \epsilon^{i_1 j_1}
           \epsilon^{i_2 j_2} \epsilon^{i_3 k_1} \epsilon^{i_4 k_2}
           \epsilon^{j_3 k_3} \epsilon^{j_4 k_4}. \nonumber
\end{eqnarray}

Using the matching relations for the $SU(6)$ scale, $\Lambda$, and the scales
of the $SU(3)$ groups $\Lambda_1^6=\Lambda_2^6=\frac{\Lambda^{12}}{v^6}$
we obtain the $SU(6)$ curve in the large VEV limit:
\begin{eqnarray}
  y^2&=& \Big[ x^3-\frac{1}{16 v^4} \Big( \frac{9}{2} (T^2) +\frac{11}{3} S^4 -S U \Big)  x
         -2 \frac{\Lambda^{12}}{v^6}- \nonumber \\
     &&   \frac{1}{64 v^6} \Big( 2 S^3 U -\frac{107}{27} S^6 - \frac{1}{4} U^2 
          - \frac{3}{2} S^2 (T^2) + 9 (T^3) \Big) \Big]^2 - 4 \frac{\Lambda^{24}}{v^{12}}.
\end{eqnarray}
After rescaling $x\rightarrow x/(4 v^2)$ and $y\rightarrow y/(64 v^6)$ the curve takes the form
\begin{eqnarray}
   y^2&=& \Big[ x^3-  \Big( \frac{9}{2} (T^2) +\frac{11}{3} S^4 -S U \Big)  x
         - 2^7 \Lambda^{12}- \nonumber \\
     &&    \Big( 2 S^3 U -\frac{107}{27} S^6 - \frac{1}{4} U^2 
          - \frac{3}{2} S^2 (T^2) + 9 (T^3) \Big) \Big]^2 - 2^{14} \Lambda^{24}.
\end{eqnarray}
As before, this is the final form of the full $SU(6)$ 
curve because any modification allowed by the global symmetries
is of the same order in $1/v$ as the terms already present.

\end{document}